\documentclass[%
reprint,
superscriptaddress,
amsmath,amssymb,
aps,
twocolumn
]{revtex4-2}

\usepackage{graphicx}
\usepackage{amssymb}
\usepackage{longtable}
\usepackage{rotating}
\usepackage{array}
\usepackage{multirow}
\usepackage{makecell}
\usepackage{booktabs}
\usepackage{xcolor}
\usepackage{ulem}
\usepackage{colortbl}
\usepackage{tikz}
\usepackage{amsmath,esint}
\usepackage[flushleft]{threeparttable}
\usepackage{hyperref}

\setcitestyle{super}

\usepackage{dcolumn}
\usepackage{bm}
\usepackage{physics}
\usepackage{hyperref}

\begin{document}

\preprint{APS/123-QED}

\title{
In-plane optically tunable magnetic states in 2D materials via tailored femtosecond laser driving
}

\author{Shuang Liu}
\affiliation{Guangdong Technion – Israel Institute of Technology, Guangdong 515063, China}
\affiliation{Department of Physics, Technion – Israel Institute of Technology, 32000 Haifa, Israel}
\author{Oren Cohen}
\affiliation{Guangdong Technion – Israel Institute of Technology, Guangdong 515063, China}
\affiliation{Department of Physics, Technion – Israel Institute of Technology, 32000 Haifa, Israel}
\author{Peng Chen}\email{peng.chen@gtiit.edu.cn}
\affiliation{Guangdong Technion – Israel Institute of Technology, Guangdong 515063, China}
\affiliation{Guangdong Provincial Key Laboratory of Materials and Technologies for Energy Conversion, Guangdong Technion -- Israel Institute of Technology, Guangdong 515063, China}
\affiliation{Department of Physics, Technion – Israel Institute of Technology, 32000 Haifa, Israel}
\author{Ofer Neufeld}\email{ofern@technion.ac.il}
\affiliation{Schulich Faculty of Chemistry, Technion – Israel Institute of Technology, 32000 Haifa, Israel}

\date{\today}

\begin{abstract}
It is well established that light can control magnetism in matter, e.g. via the inverse Faraday effect or ultrafast demagnetization. However, such control is typically limited to magnetization transverse to light’s polarization plane, or out-of-plane magnetism in 2D materials, while in-plane magnetic moments have remained largely unexplored. This is due to the difficulty of generating electronic orbital angular momentum components within light's polarization plane. 
Here we overcome this limitation, demonstrating complete three-dimensional, all-optical control of magnetism in 2D materials. Using first-principles simulations, we show that a tailored, two-color laser field can induce and steer magnetic moments in any direction with the relative angle between the laser polarizations playing a key parameter in coherent control.
We analyze the physical mechanism of this process and show that it arises from a simultaneous breaking of time-reversal and spatial-inversion symmetries in the two-color laser. In-plane orbital moments are introduced via non-zero out-of-plane longitudinal photogalvanic currents enabled by broken inversion and mirror symmetries, while time-reversal symmetry breaking enables build-up of spin-rotation processes through spin-orbit coupling, translating the orbital moments to transient magnetism.
Our findings demonstrate a full 3D coherent control scheme for transient magnetic states on femtosecond timescales driven by tailored lasers, and can be used to develop novel spectroscopies for magnetism, all-optical magnetic switching for ultrafast spintronics, and novel information storage capabilities.

\end{abstract}

\maketitle
 
Magnetism is a fundamental physical phenomenon with far-reaching technological applications in information storage and processing\cite{Magnetism_device_Bhatti2017}. One key challenge within this field is pushing operational speeds of devices to faster timescales, ideally down to the femtosecond regime\cite{Ultrafast_magnetism_Marini2022,Ultrafst_magnetism_Kampfrath2023,Ultrafast_magnetism_Fang2024,Ultrafast_magnetism_Siegrist2019,PHz_electronics_Heide2024}. Such ultrafast control is not possible with conventional approaches such as external magnetic field, but over the past two decades has been explored and improved via laser-induced coherent control of magnetic states\cite{laser_magnetism_Kimel2007,laser_magnetism_Kalashnikova2015,laser_magnetism_Kirilyuk2010,laser_magnetism_Koopmans2000,laser_magnetism_Li2022,laser_magnetism_Walowski2016}.
Intense laser pulses irradiated onto a magnetic material have unlocked a wealth of ultrafast magnetic phenomena, including femtosecond demagnetization induced by infra-red\cite{laser_demagnetization_Elliott2016,laser_demagnetization_Scheid2021} and X-ray pulses\cite{Xray_exp_Cardin2020,Xray_exp_Chen2018,Xray_exp_Lveill2022,Xray_exp_Schneider2020,Xray_theo_Kapcia2022,Xray_theo_Kapcia2023,Xray_theo_Kapcia2024}, spin transfer\cite{laser_spin_transfer_Hofherr2020,laser_spin_transfer_Steil2020,laser_spin_transfer_Tengdin2020,laser_spin_transfer_Willems2020,laser_spin_transfer_Schellekens2014,laser_spin_transfer_Siegrist2019}, spin switching\cite{laser_spinflip_Stanciu2007}, and magnetic phase transitions\cite{laser_AFM2FM_Golias2021,laser_AFM2FiM_He2023}.
It was also shown that magnetization can be optically induced in initially non-magnetic materials via various mechanisms\cite{laser_NM2FM_He2022,laser_NM2FM_Neufeld2023,Laser_NM2FM_Okyay2020,Laser_NM2FM_Marini2022,Laser_NM2FM_Elliott2016,nonlinear_Edelstein_ik2024,cep_neufeld2025,laser_magnetsim_sangeeta2025}. The ability to optically induce magnetism represents a fundamental shift, offering a pathway to engineer magnetic properties in materials not traditionally considered magnetic.

In addition, two-dimensional (2D) materials are an ideal platform for realizing ultrafast magnetic control, owing to their strong potential for device integration via heterostructure engineering and their superior controllability compared with bulk materials.\cite{2D_magnetism_Deng2018,2D_magnetism_He2020,2D_magnetism_OHara2018,2Dmagnet_Gong2017,2Dmagnet_Huang2017}.
However, all previous studies in 2D materials were limited to tuning out-of-plane magnetization, parallel to light's propagation axis. 
Indeed, in-plane magnetism provides additional degrees of freedom besides the magnitude of the magnetic moment - it's direction, which has potential for non-boolean logic operations\cite{non_Boolean_Datta2012}. Therefore, attaining full 3D control of ultrafast transient magnetic states is a crucial next step, promising to next-generation spintronic devices. 

In parallel to advancements in ultrafast magnetism, strong-field physics was introduced to condensed matter\cite{Strong_field_Baltuka2003,Strong_field_Cavaletto2024,Strong_field_Ghimire2014,Strong_field_Goulielmakis2007,Strong_field_Hui2021,Strong_field_Schultze2012,Strong_field_Uzan2020} with rich applications such as tailoring valley pseudo-spin \cite{SF_valley_JimnezGaln2021,Laser_topology_JimnezGaln2021,Ultrafast_valley_switching_Rana2023} and valley currents\cite{laser_valley_current_Sharma2023}, steering Dirac electrons\cite{SF_Dirac_current_Reimann2018}, high harmonic generation (HHG)\cite{HHG_Bai2020,HHG_Schmid2021,2color_HHG_Kim2005,2color_HHG_Qiao2019,2color_HHG_Frolov2010,2color_HHG_Cormier2000} and spin current generation\cite{SF_spin_current_Gill2025}. In particular, bichromatic two-color intense fields have been employed for enhanced tunability in light-matter interactions due to their ability to break additional symmetries and provide multiple control knobs such as relative phases, polarizations, intensities etc.\cite{tailored_laser_Hamilton2017,2color_Eickhoff2021,2color_HHG_Navarrete2020,2color_HHG_Li2017,2color_HHG_Song2020,2color_phase_Neufeld2025,2color_Habibovi2024}.
While these tailored fields have been successfully used to steer electrons and generate high-order harmonics, their potential to control magnetism has remained untapped. This presents a compelling opportunity --- to apply the precision of bichromatic fields to the challenge of ultrafast magnetization dynamics.


Here, we demonstrate 3D optical control of magnetism by investigating the response of non-magnetic 2D materials to intense two-color laser pulses, using state-of-the-art time-dependent spin-density functional theory (TDSDFT).
We show that when a second in-plane linearly-polarized pulse at frequency $\omega_2 = 2\omega_1$ is applied in tandem to a circularly polarized pulse at frequency $\omega_1$, a stable in-plane magnetism will be induced.
We investigate the physical mechanism behind this effect and show that the breaking of mirror symmetry along the \textit{z}-axis (transverse to the monolayer plane) plays a crucial role. Unlike earlier works that only involved angular momentum along the \textit{z}-axis, our scheme enables generation of in-plane components that arise from a longitudinal photogalvanic effect along the z-axis (which becomes allowed with the broken symmetry, we denote it as ‘photogalvanic current’ because it exhibits a zero-frequency term as shown in section VII of SI). This in-plane angular momentum translates to in-plane transient magnetic states through spin-orbit interactions.
Remarkably, we find that bichromatic two-color fields provide a simple coherent control knob for rotating the induced magnetic state in the monolayer plane in the form of the relative polarization angle between the two pulses. 
Our results  highlights the potential of tailored laser fields as powerful tools for full 3D optical manipulation of magnetism, crucial for advancing the design of ultrafast all-optical spintronics. 

\begin{figure}[htp]
    \centering
    \includegraphics[width=1.0\linewidth]{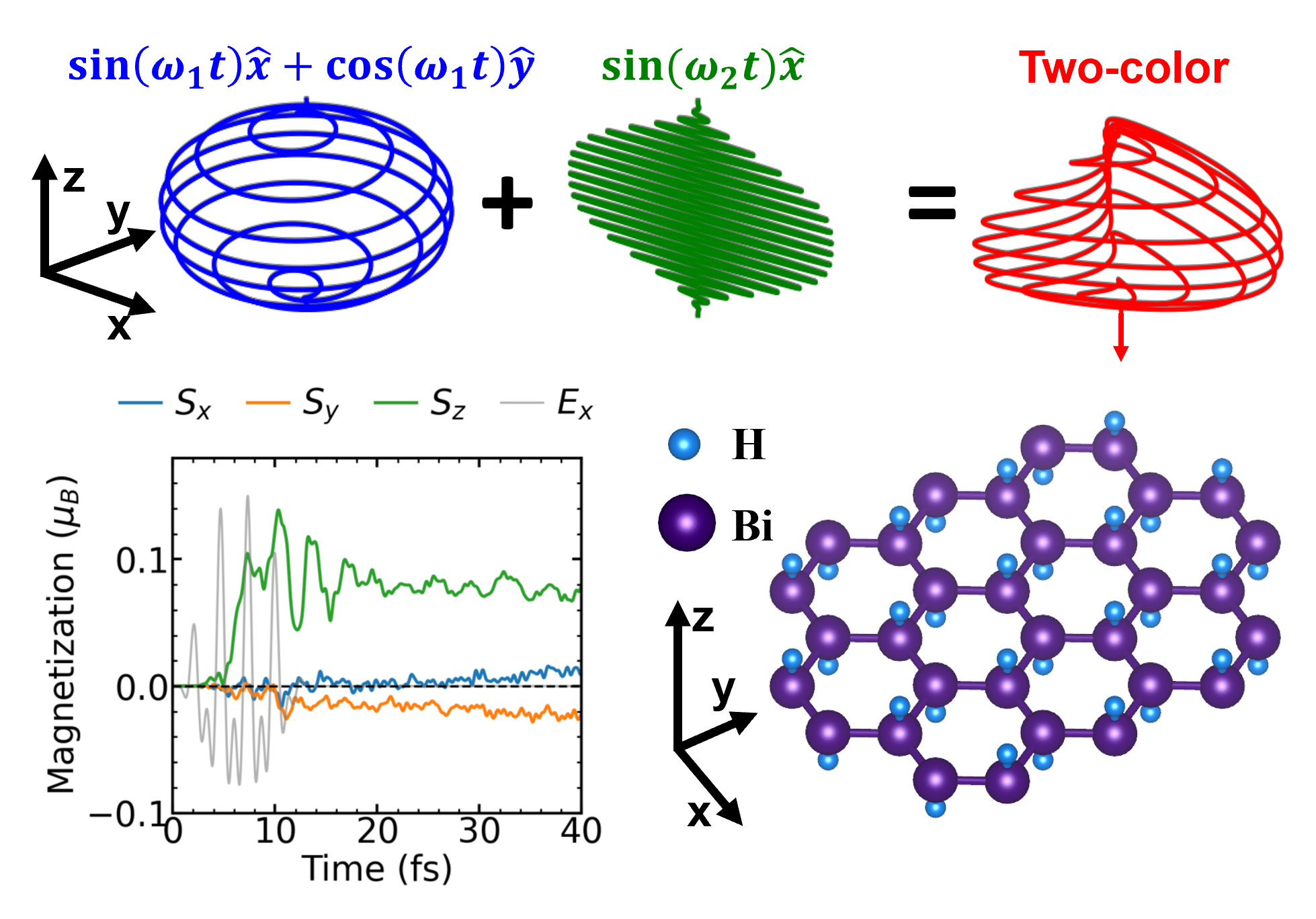}
    \caption{Magnetism dynamics induced by two-color field in BiH. (a) Illustration of bichromatic laser field: a circularly polarized laser with frequency $\omega_1$ and a linearly polarized laser with frequency $\omega_2$ are superimposed, $\omega_2 = 2\omega_1$, with $\omega_1$ corresponding to 400 nm light. (b) structure of BiH, where purple atoms represent bismuth and blue atoms represent hydrogen. (c) Calculated induced magnetism in BiH induced by the two-color field. The external field is applied over a duration of 8.34 fs. It comprises a two-color pulse combining a circularly polarized fundamental field ($\omega_1$) and a linearly polarized second harmonic ($\omega_2 = 2\omega_1$) oriented along the $x$-axis. Laser field is set to an intensity of $10^{11}~\mathrm{W/m^2}$, with relative strength parameters $\delta_1 = \delta_2 = 1$.}\label{figs:overview}
\end{figure}

Let us begin by describing our theoretical approach. To model light-induced magnetization dynamics, we employ \textit{ab-initio} calculations based on time-dependent spin-density functional theory in the Kohn–Sham (KS) formulation, which is well-established for ultrafast magnetic phenomena such as laser-induced demagnetization. The ground state is obtained from spin-polarized DFT and then propagated in real-time using the following equations of motion:
\begin{equation}
i \partial_t \ket{\psi^{\mathrm{KS}}_{n,\mathbf{k}}(t)} = \left[ \frac{1}{2} \left(-i \nabla + \frac{\mathbf{A}(t)}{c} \right)^2 \sigma_0 + \mathrm{v}_{\mathrm{KS}}(t) \right] \ket{\psi^{\mathrm{KS}}_{n,\mathbf{k}}(t)}
\end{equation}
where $\ket{\psi^{\mathrm{KS}}_{n,\mathbf{k}}(t)}$ is the spinor Bloch state at $k$-point $\mathbf{k}$ and band index $n$, $\sigma_0$ is the $2 \times 2$ identity matrix, $\mathrm{v}_{\mathrm{KS}}(t)$ is the time-dependent KS potential (see more details on ref\cite{laser_NM2FM_Neufeld2023}), and $\mathbf{A}(t)$ is the vector potential of the laser field (within the dipole approximation, such that $-\partial_t \mathbf{A}(t) = c \mathbf{E}(t)$, where $c \approx 137.036$ is the speed of light in atomic units). 

The KS equations are solved in a real-space grid representation with Octopus code\cite{Octopus_TancogneDejean2020}. From the time-propagated states we calculate the time-dependent current, $\mathbf{J}(t)=\frac{1}{\Omega} \int_{\Omega} d^{3} r \mathbf{j}(\mathbf{r}, t)$, where $\Omega$ is unit cell volume and $\mathbf{j}(\mathbf{r}, t)$ is the microscopic time-dependent current density:

\begin{equation}
\mathbf{j}(\mathbf{r}, t)=\sum_{n, \mathbf{k}, a}\left[\begin{array}{c}
\varphi_{n, \mathbf{k}, a}^{\mathrm{KS}\;*}(\mathbf{r}, t)(
\frac{1}{2}(-i \nabla+\frac{\mathbf{A}(t)}{c})\\ 
+[V_{ion}, \mathbf{r}]) \varphi_{n, \mathbf{k}, a}^{\mathrm{K S}}(\mathbf{r}, t)
+c.c.\end{array}\right]
+\mathbf{j}_{m}(\mathbf{r}, t)
\end{equation}

where $\mathbf{j}_{m}(\mathbf{r}, t)$ is the magnetization current density (which after spatial integration vanishes). The spin expectation values are calculated as $\langle\mathbf{S}(t)\rangle=$ $\left\langle\psi_{n, \mathbf{k}}^{K S}(t)\right| \mathbf{\hat{S}}\left|\psi_{n, \mathbf{k}}^{K S}(t)\right\rangle$, and are used to track the light-induced magnetization after spatial integration, and more technical details on the propagation scheme and numerical parameters can be seen in ref\cite{laser_NM2FM_Neufeld2023} and in the supplementary information (SI).

The interaction with the laser field is included in the velocity gauge\cite{Octopus_Andrade2015} using the vector potential:
\begin{equation}
\mathbf{A}(t) = f(t) [\frac{c\delta_1 E_0}{\omega_1} \sin(\omega_1 t) \hat{e}_{1}+\frac{c\delta_2 E_0}{\omega_2} \sin(\omega_2 t+\Delta\phi) \hat{e}_{2}]
\end{equation}
where $f(t)$ is the pulse envelope that is taken to be ``super-sine'' pulse\cite{supersine_Neufeld2019} (for details see the SI).
$E_0$ is the peak field amplitude and $\delta_1$ and $\delta_2$ are the relative strength. $\omega_1$ and $\omega_2$ are the carrier frequency, and $\hat{e}_{1}$ and $\hat{e}_{2}$ are unit vectors defining the polarization direction. 
As illustrated in Fig. 1(a), the two-color field is constructed by combining a circularly polarized ($\hat{e}_{1}=\frac{1}{\sqrt2}(\hat{e_x}+i\hat{e_y})$) laser of fundamental frequency $\omega_1$, with a second linearly polarized laser field ($\hat{e}_{2}$) of frequency $\omega_2$. In the exemplary case shown in Fig. 1(a), we have $\hat{e}_{2}=\hat{e_x}$ and $\omega_2=2\omega_1$ with $\Delta\phi=0$. This combination exhibits a broken inversion symmetry\cite{Floquet_throry_Neufeld2019}. The resulting asymmetry in $\mathbf{A}(t)$ should modify nonlinear optical effects, including the magnetic response.
In our main simulations, this two-color field irradiates a monolayer of bismuth hydride (BiH)\cite{BiH_Song2014}, as shown in Fig. 1(b), which is an ideal platform for exploring laser-induced magnetism in 2D materials due to strong spin–orbit coupling (SOC) from the heavy bismuth atoms\cite{laser_NM2FM_Neufeld2023}.

Fig. 1(c) present exemplary light-driven magnetization dynamics. As expected, the laser pulse induces an out-of-plane spin polarization ($S_{z}$) that reaches approximately 0.07 µB (green curve), a result consistent with previous reports\cite{laser_NM2FM_Neufeld2023}. Surprisingly, in addition to $S_z$, a significant and persistent in-plane spin component along the $y$-axis ($S_y$), ~$0.02 \mu_B$, is also observed (see orange curve), which is absent when driving with pure circular monochromatic laser ($\delta_2=0$ in eq. 3). The $x$-component of spin ($S_x$) remains nearly zero throughout the system's evolution.
We note that although the incident laser does not carry a total in‑plane angular momentum, it breaks inversion and rotational symmetry of BiH. As a result, conservation of in-plane angular momentum is irrelevant, permitting in-plane magnetism to arise, and allowing for potential tuning of the induced magnetism via different laser fields by changing their parameters.

\begin{figure}[htp]
    \centering
    \includegraphics[width=1.0\linewidth]{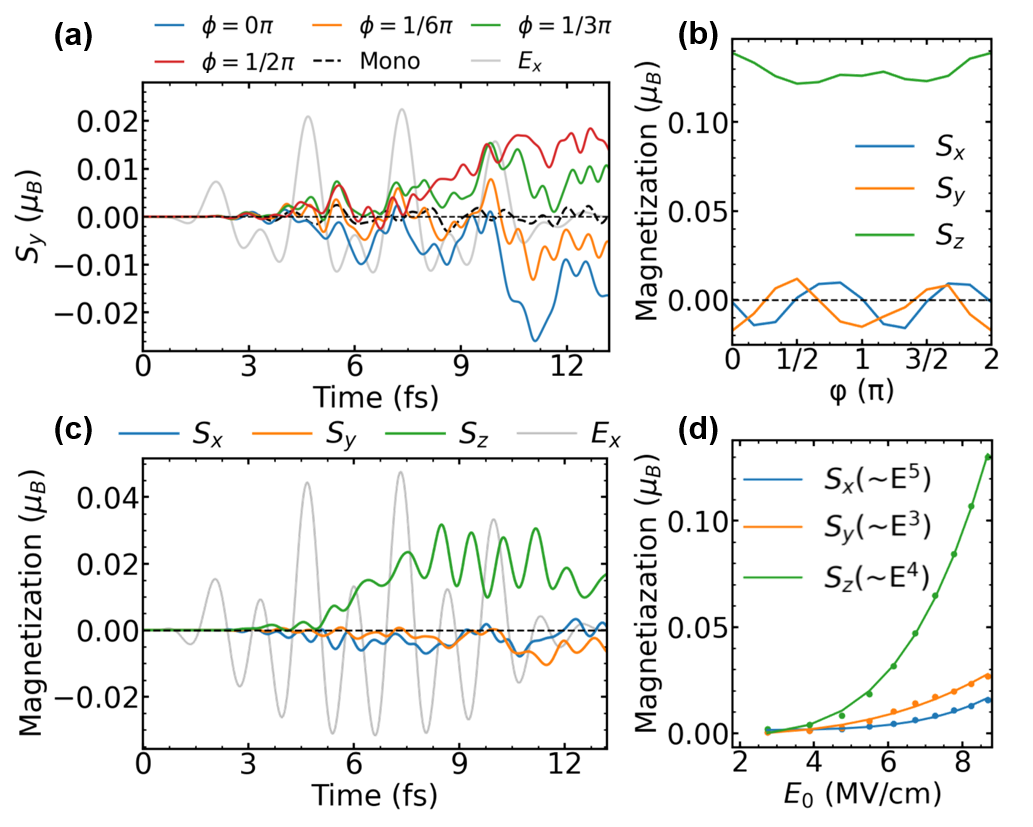}
    \caption{Coherent control of femtosecond laser-induced magnetism in BiH. (a) Dynamics of $S_y$ as a function of the relative polarization angle $\varphi$. The simulation employs a 6.67 fs two-color pulse ($\omega_2 = 2\omega_1$) with an intensity of $10^{11}~\mathrm{W/m^2}$. The relative strength parameters are $\delta_1 = \delta_2 = 1$, and the polarization vector of the second harmonic is defined as $\hat{\mathbf{e}}_{2}=(\cos\varphi \sin\theta, \sin\varphi \sin\theta, \cos\theta)$). (b) Dependence of net induced magnetism on $\varphi$. (c) Dynamics of spin
    under two-color pulse ($\omega_2 = 2\omega_1$) with a duration of 6.67 fs, the linearly polarized second harmonic oriented along the $x$-axis.  The laser intensity is set to $10^{11}~\mathrm{W/m^2}$, with relative strength parameters $\delta_1 = 0.4$ and $\delta_2 = 1$. (d) Relation between induced magnetism and driving laser electric field amplitude indicating a highly nonlinear process.}\label{figs:Modulation}
\end{figure}

Having verified that both in-plane and out-of-plane magnetism can be induced by the bi-chromatic light, we move on to explore coherent control of this induced magnetism by tuning the two-color laser field parameters. Specifically, we have systematically explored different polarization directions, relative frequencies, intensities, and relative phases, within the two-color drive.

We find that the relative phase ($\Delta\phi$) between the two laser fields exhibits minimal influence on the magnetic response (for details see Fig. S1 in SI). Notably, in ultrafast and nonlinear optics in solids this relative phase is typically a strong control parameter\cite{2color_phase_Rana2022,2color_phase_Neufeld2025,2color_phase_Sanchez2023,2color_phase_Tyulnev2024,2color_phase_Sederberg2020,2color_phase_https://doi.org/10.48550/arxiv.2507.05768}; however, our results indicate that it likely plays a minimal physical role in nonlinear magnetization dynamics. Similarly, changing the relative frequencies to $\omega _2=n\omega_1$ for $n=3,4,5,6$ is found to reduce the out-of-plane magnetism. Especially, the induced $S_y$ component changes sign for $n=3$ (see Fig. S2 in SI).

In contrast, the relative polarization angle of the $2\omega$ beam (denoted $\theta$ and $\varphi$, with the polarization vector $\hat{e}_{2}=(\mathrm{cos}\varphi \mathrm{sin}\theta, \mathrm{sin}\varphi \mathrm{sin}\theta, \mathrm{cos}\theta )$) and intensity of the two-color laser are found to have a very pronounced impact on the direction and magnitude of the induced magnetization. Figure 2 explores this effect, while fixing the circularly-polarized laser parameters and varying the polarization direction of the linearly polarized $2\omega$ pulse. Initially, we fixed the out-of-plane angle at $\theta=0$, ensuring that the linear polarization remains entirely in-plane, and varied the in-plane polarization angle $\varphi$. Figure 2(a) shows the $S_y$ magnetization temporal dynamics under different $\varphi$ - as the $2\omega$ pulse in-plane polarization direction varies, the induced $S_y$ component is tuned all the way from negative to positive values, while it vanishes if purely a monochromatic circularly polarized laser drives the system. The amplitude of $S_y$ reaches a maximum of 0.02 $\mu_B$, up to ~20\% of the out-of-plane component, indicating a non-negligible in-plane induced magnetic state. By further reducing the amplitude of the circular $\omega_1$ component (reducing the level of time-reversal symmetry breaking), the ratios between in-plane and out-of-plane magnetization can be further controlled and enhanced up to 40\% (Fig. 2(c)).

Figure 2(b) shows the induced magnetization components after the laser pulse ends, demonstrating an expected 360$^\circ$ periodic dependence on $\varphi$. Interestingly, a clear $S_x$ component simultaneously arises and oscillates in accordance with $S_y$, showing that by tuning $\varphi$ the in-plane magnetization can be coherently and precisely rotated. 
Both in-plane magnetization components, $S_x$ and $S_y$, complete two full oscillation cycles as $\varphi$ progresses through a single 360° rotation (see further discussion in section IV, Fig. S3 in SI). 

The magnetization dynamics under different $\theta$ is shown in Fig. S4, demonstrating that as $\theta$ increases — corresponding to a gradual rotation of the linear polarization axis out of the plane — the in-plane magnetization gradually decreases in magnitude. This highlights the role of breaking inversion symmetry while still driving electrons directly in-plane for generation of orbital angular momentum. 


Another laser parameter that significantly influences the induced magnetism is its intensity, which is proportional to the square of the field amplitude, $E_0$. Fig. 2(d) shows that increasing $E_0$ leads to a corresponding increase in the in-plane magnetization amplitude. By fitting the data to a power-law relation, we find a prominent nonlinear dependence for all magnetization components, proving that multi-photon transitions are involved (as expected in this nonlinear regime\cite{cep_neufeld2025}). The laser intensity therefore serves as an efficient and tunable parameter for controlling the magnetization amplitude.

\begin{figure}[htp]
    \centering
    \includegraphics[width=1.0\linewidth]{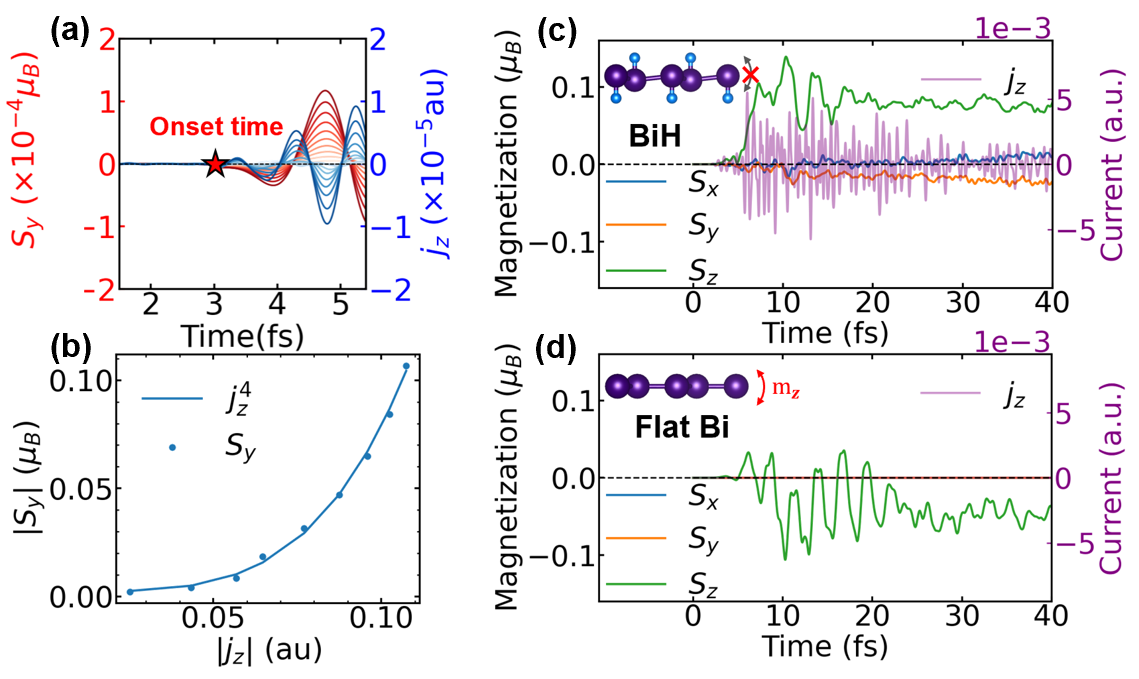}
    \caption{Physical mechanism of induced in-plane magnetism. (a) Femtosecond dynamics of out-of-plane photocurrents, $j_z(t)$, labeled with blue and $S_y(t)$ labeled with red, deeper color indicates stronger laser intensity ($10^{10}\sim10^{11}$ W/m$^2$), other parameters correspond to those in Fig. 1 (c). (b) Relation between amplitude of laser-induced $j_z$ and $S_y$ terms. (c) and (d) Induced magnetization dynamics and $j_z(t)$ (laser parameters correspond to those in in Fig. 1 (c)) in (c) BiH (mirror asymmetric) and (d) pristine flat bismuth (mirror symmetric).}\label{figs:Mechanism}
\end{figure}

The above results clearly demonstrate that bichromatic driving not only induces in-plane magnetism, but also freely controls the direction and amplitude of both in-plane and out-of-plane magnetization components. 

We now turn to explore the physical mechanism responsible for this effect.
Since the nonlinear magnetization in our regime arises from spin-orbit coupling (in the absence of external magnetic fields this is the only mechanism that can cause ultrafast spin flip processes), the key to generate in-plane magnetism lies in the emergence of a non-zero out-of-plane current $j_z(t)$ --- this is perquisite for in-plane electronic orbital angular momentum, which gives effective magnetic torque on the spin-rotations within the monolayer plane. Essentially, for the angular momentum vector to have $x$ or $y$ components, the electronic excitation must have an out-of-plane $z$ component due to the mathematical structure of angular momentum. Figure 3(a) shows the temporal light-induced dynamics of $S_y$ and $j_z$ for different laser intensities. This temporal evolution reveals that their onset times coincide for all driving intensities, indicating that both are excited simultaneously. Furthermore, we find that $S_y$ and $j_z$ show a strongly-correlated (power-law) increase in amplitude as the laser intensity varies. These results highlight that $j_z$ is not only prerequisite for the occurrence of $S_y$, but is directly involved in its generation mechanism, in accordance with angular momentum considerations.

\begin{figure}[htp]
    \centering
    \includegraphics[width=1.0\linewidth]{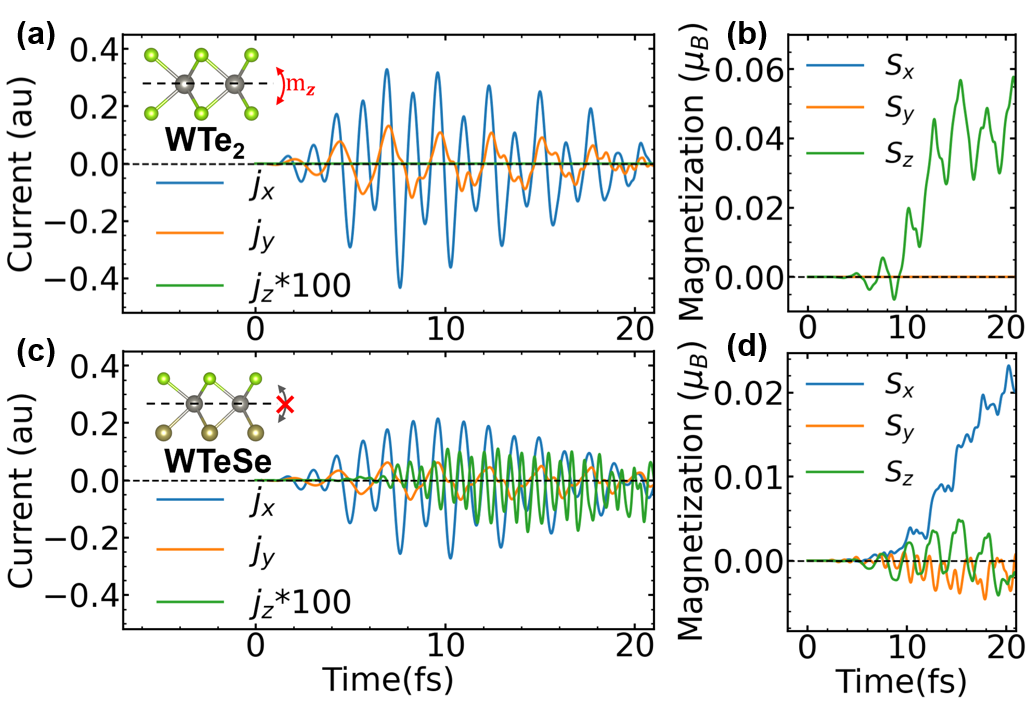}
    \caption{Induced in-plane magnetism in Transition metal dichalcogenides (TMDs) and Janus TMDs. Induced photocurrents (laser parameters correspond to those in Fig. 3(c) except the laser duration time is 10 fs) in (a) WTe$_2$ and (c) WSeTe (Janus TMD). Induced magnetism dynamics in (b) WTe$_2$ and (s) WSeTe.}\label{figs:TMDs}
\end{figure}

Notably, a nonzero out-of-plane current is allowed in our set-up even though BiH is inversion symmetric because: (i) BiH is not mirror-symmetric along the z-axis (see typical current evolution in Fig. 3(c));  and (ii) the two-color laser field breaks inversion symmetry and circular component\cite{photogalvanic_Neufeld2021,Floquet_throry_Neufeld2019}.

To validate this physical mechanism and verify the crucial role of mirror symmetry breaking, we performed similar calculations on pristine monolayer bismuth (Bi), which adopts a planar (graphene-like) structure that preserves mirror symmetry (see Fig. 3(d)). As expected, the induced $j_z$ remains zero (forbidden), and accordingly, no in-plane magnetism is observed even though a similar two-color drive is employed.
To further confirm that it is the breaking of mirror symmetry, rather than simply the presence of hydrogen atoms with contributed $sp^3$-hybridized conduction electrons that might lead to $j_z$ formation, we examined a buckled bismuth structure without hydrogen (see Fig. S5 of SI). Although the induced out-of-plane magnetization remains similar to the pristine case, a non-zero $j_z$ is now generated, resulting in the excitation of in-plane magnetism as well. 

While broken mirror symmetry is a prerequisite for a non-zero $j_z$, the circular laser pulse is equally critical, as linear driving imposes additional symmetry constraints that suppress the response (see Fig. S6(a)). Crucially, introducing a $2\omega$ field breaks the inversion symmetry of the Hamiltonian. This not only strongly enhances the induced $j_z$ amplitude, but also fundamentally alters its spectral composition—leading to a symmetry-allowed significant DC (zero frequency) component (see Fig. S6(b)), i.e. bulk photogalvanic injection currents. This rectified current enables the effective accumulation of in-plane magnetism as it allows non-vanishing electronic angular momenta generation (see Fig. S6(c)), which can be transferred to the spin system via SOC. This highlights the essential interplay between structural mirror symmetry breaking and the tailored $\omega-2\omega$ laser field in establishing 3D magnetic control (see details in section VII of SI).


Lastly, we extended our analysis to other material systems. Transition metal dichalcogenides (TMDs) containing heavy elements such as tungsten (W), rhenium (Re), and tantalum (Ta) are promising candidates due to their strong spin–orbit coupling and layered structures\cite{TMD_Kadek2023}. We explored WSe$_2$ and WSeTe as representative TMDs to complement our results and gain deeper insight into the underlying physical mechanisms.
As shown in Fig. 4(a, b), WSe$_2$, which possesses mirror symmetry, exhibits zero $j_z$ and purely out-of-plane light-induced magnetism. In contrast, WSeTe, which is a Janus TMD\cite{2D_janus_Yagmurcukardes2020,WSeTe_Kumar2024} and breaks mirror symmetry, displays a non-zero $j_z$ and complementary induced in-plane magnetism (see Fig. 4(c, d)). Compared to BiH, TMDs are more experimentally accessible and well-studied, thereby offering a practical platform to verify and apply our theoretical predictions. These results strongly support the generality and robustness of this proposed physical mechanism for generating in-plane magnetism, particularly the crucial roles of breaking mirror, inversion, and time-reversal symmetries in enabling in-plane magnetism.

To summarize, we have theoretically predicted with state-of-the-art \textit{ab-initio} simulations the emergence of stable in-plane transient magnetism in 2D materials induced by coherent two‑color ($\omega-2\omega$) laser fields. Our simulations reveal that a non‑zero photocurrent along the $z$‑direction (light's propagation axis), originating from a combination of broken mirror symmetry along $z$-direction and broken inversion symmetry in the bichromatic drive, is the essential mechanism for driving the generation of in-plane magnetism. This longitudinal photogalvanic current component enables in-plane electronic orbital currents, which lead to induced magnetism via spin-orbit interactions.
Crucially, we showed that the relative polarization angle between the $\omega$ and $2\omega$ components of the drive serves as a powerful control parameter, enabling precise tuning of both the orientation and magnitude of the induced magnetization on ultrafast timescales, with an amplitude and in-plane/out-of-plane ratio that can be further controlled with the intensity of the laser components. We demonstrated the generality of this result and transferability of the physical mechanism with equivalent simulations in TMDs.

In contrast to out-of-plane magnetism (which stores a bit as “up/down”), in-plane magnetization allows a full 360$^\circ$ of orientation. Notably, even though we show that a mirror-symmetry broken material is a requirement for obtaining in-plane magnetism, this is readily achieved anyways in many encapsulation schemes\cite{Asymmetry_stack_Zhou2021,Asymmetry_stack_https://doi.org/10.48550/arxiv.2503.17947}.
Consequently, we envision that the orientation of in-plane magnetization can act as a continuous or multi-level variable for encoding data with multi-state logic, and paving the way to non‑Boolean operations, or achieving multilevel memory and vector logic within single spintronic devices. Our results show that this can be achieved on ultrafast timescales, compatible with Petahertz electronics and spintronics\cite{PHz_electronics_Heide2024}.
\\
\noindent {\bf Acknowledgment}\\
The work was supported by the National Natural Science Foundation of China (Grant No. 12574092). ON gratefully acknowledges the scientific support of Prof. Dr. Angel Rubio and the Young Faculty Award from the National Quantum Science and Technology program of Israel's Council of Higher Education Planning and Budgeting Committee.  \\


\end{document}